\documentclass[10pt,letterpaper]{article}

\usepackage[numbers]{natbib}
\usepackage{amsmath} 
\usepackage[version=4]{mhchem}	
\DeclareMathOperator{\tr}{tr}
\usepackage{graphicx}
\usepackage[export]{adjustbox}
\usepackage[utf8]{inputenc}
\usepackage{nameref,hyperref}
\usepackage[left]{lineno}
\usepackage{microtype}
\DisableLigatures[f]{encoding = *, family = * }
\usepackage{setspace} 
\usepackage{changepage}
\usepackage[aboveskip=1pt,labelfont=bf,labelsep=period,singlelinecheck=off]{caption}
\makeatletter
\renewcommand{\@biblabel}[1]{\quad#1.}
\makeatother
\usepackage{lastpage,fancyhdr,graphicx}
\usepackage{epstopdf}
\usepackage{color}
\definecolor{Gray}{gray}{.25}
\usepackage{graphicx}
\usepackage{sidecap}
\usepackage{wrapfig}
\usepackage[pscoord]{eso-pic}
\usepackage[fulladjust]{marginnote}
\reversemarginpar

\begin{document}
\vspace*{0.35in}

\begin{flushleft}
{\Large
\textbf\newline{Imaging linear and circular polarization features in leaves with complete Mueller matrix polarimetry}
}
\newline
\\
C.H. Lucas Patty\textsuperscript{1*},
David A. Luo\textsuperscript{2},
Frans Snik\textsuperscript{3},
Freek Ariese\textsuperscript{4},
Wybren Jan Buma\textsuperscript{5},
Inge Loes ten Kate\textsuperscript{6},
Rob J.M. van Spanning\textsuperscript{7},
William B. Sparks\textsuperscript{8},
Thomas A. Germer\textsuperscript{9},
Gy\H oz\H o Garab\textsuperscript{10,11},
Michael W. Kudenov\textsuperscript{2}
\\
\bigskip
\tiny 1 Molecular Cell Physiology, VU Amsterdam, De Boelelaan 1108, 1081 HZ Amsterdam, The Netherlands
\\
{2} Optical Sensing Lab, Department of Electrical and Computer Engineering, North Carolina State University, Raleigh, NC 27695, USA
\\
{3} Leiden Observatory, Leiden University, P.O. Box 9513, 2300 RA Leiden, The Netherlands
\\
{4} LaserLaB, VU Amsterdam, De Boelelaan 1083, 1081 HV Amsterdam, The Netherlands
\\
{5} HIMS, Photonics group, University of Amsterdam, Science Park 904, 1098 XH Amsterdam, The Netherlands
\\
{6} Department of Earth Sciences, Utrecht University, Budapestlaan 4, 3584 CD Utrecht, The Netherlands
\\
{7} Systems Bioinformatics, VU Amsterdam, De Boelelaan 1108, 1081 HZ Amsterdam, The Netherlands
\\
{8} Space Telescope Science Institute, 3700 San Martin Drive, Baltimore, MD 21218, USA
\\
{9} Senior Science Division, National Institute of Standards and Technology, 100 Bureau Drive, Gaithersburg, MD 
20899, USA \\
{10} Institute of Plant Biology, Biological Research Centre, Hungarian Academy of Sciences, P.O. Box 521, H-6701 Szeged, Hungary\\
{11} Department of Physics, Faculty of Science, University of Ostrava, Chittussiho 10, Slezsk\' a Ostrava, Czech Republic

\bigskip

\bf *lucas.patty@vu.nl

\end{flushleft}

\section*{Abstract}
Spectropolarimetry of intact plant leaves allows to probe the molecular architecture of vegetation photosynthesis in a non-invasive and non-destructive way and, as such, can offer a wealth of physiological information. In addition to the molecular signals due to the photosynthetic machinery, the cell structure and its arrangement within a leaf can create and modify polarization signals. Using Mueller matrix polarimetry with rotating retarder modulation, we have visualized spatial variations in polarization in transmission around the chlorophyll \textit{a} absorbance band from 650 nm to 710 nm. We show linear and circular polarization measurements of maple leaves and cultivated maize leaves and discuss the corresponding Mueller matrices and the Mueller matrix decompositions, which show distinct features in diattenuation, polarizance, retardance and depolarization. Importantly, while normal leaf tissue shows a typical split signal with both a negative and a positive peak in the induced fractional circular polarization and circular dichroism, the signals close to the veins only display a negative band. The results are similar to the negative band as reported earlier for single macrodomains. We discuss the possible role of the chloroplast orientation around the veins as a cause of this phenomenon. Systematic artefacts are ruled out as three independent measurements by different instruments gave similar results. These results provide better insight into circular polarization measurements on whole leaves and options for vegetation remote sensing using circular polarization.

\section*{Keywords}
Photosynthesis, Mueller matrix polarimetry, circular dichroism, chloroplast, chlorophyll a
\begin{center}

\end{center}


\section*{Introduction}
One of the most distinctive and characteristic features of life is the homochirality of its molecular building blocks \cite{Macdermott1996}. Chiral molecules in their most simple form exist in left-handed (L-) and a right-handed (D-) versions, called enantiomers. In non-biological systems, the mixture is expected to be racemic ($50$ $\%$ - $50$ $\%$). However, biological systems tend to have nearly 100 \% preference for one type of enantiomer, which is a feature called homochirality. In fact, the functioning and structure of biological systems is largely determined by their chiral constituents. Although there are a few exceptions \cite{Fujii2004}, amino acids mainly occur in the L-configuration and sugars occur predominantly in the D-configuration. Apart from these small molecules, many large scale molecular architectures, dimensions of which can range over several orders of magnitude, are chiral. An example of such large-scale chirality is displayed by the DNA molecule, which is always right-handed and can be over 2 m long \cite{Urry2016}. Chirality can also be observed in the chlorophylls and bacteriochlorophylls, in particular when utilized in photosynthesis (as their intrinsic signal is very weak due to their planar and almost symmetrical structure). Additionally, these chlorophylls are organized in a supramolecular structure that itself is chiral too \cite{Garab2009}.

The molecular dissymmetry of chiral molecules has a specific response to electromagnetic radiation \cite{Fasman2013} and this response both depends on the intrinsic chirality of the molecules and on the chirality of the supramolecular architecture. Examples of available spectroscopic methods that are based on this interaction are circular diattenuation (dichroism) and linear diattenuation spectroscopy. Both methods are complementary and offer valuable insight into the functionality and structure of molecules and have a long history in the research on photosynthesis \cite{Garab2009}. In circular dichroism spectroscopy, the differential extinction of left- and right-handed circularly polarized light as a function of wavelength is measured. Linear diattenuation spectroscopy characterizes the change in extinction depending on the linear polarization of the incident (orthogonal) beams. Usually, only isolated molecules or cell constituents are measured, but it has recently been shown that the circular dichroism of whole leaves can also be determined \cite{Toth2016, Patty2017}. This is not possible in linear diattenuation spectroscopy, since the retrieval of structural information is dependent on the molecular alignment of the sample. In a randomly oriented sample, such as in a leaf, this information is therefore averaged out.

Mueller matrix polarimetry allows a thorough characterization of the polarization properties of a sample. The complete Mueller matrix is a $4 \times 4$ matrix with real elements that completely describes the polarization response of an optical element. Within its elements it additionally contains polarization properties, i.e., circular and linear diattenuation,  retardance, and depolarization. Diattenuation is similar to dichroism, although the latter is ususally described in terms of absorbance. The retardance describes the phase changes of light and is independent of the intensity transmittance. The depolarization describes the ratio of incident light that becomes unpolarized upon interaction with the sample. The mathematical descriptions of these quantities will be given below.

Both linear and circular dichroism spectroscopy depend on the modulation of the incident light and the subsequent differential interaction within the sample resulting in a measurable difference. Induced linear polarization is also measurable and scattered linear polarization has been investigated for vegetation remote sensing \cite{Peltoniemi2015, Vanderbilt1985, Vanderbilt1985a, Vanderbilt2017, Grant1993}. Although it has been suggested that linear polarization remote sensing offers no additional information compared to the scalar reflectance \cite{Peltoniemi2015}, it has recently been suggested that it could be a promising remote sensing tool for the detection of leaf structural changes such as brought upon by drought \cite{Vanderbilt2017}.

Also circular polarization by photosynthesic systems might potentially be a powerful tool for the remote sensing of biosignatures on Earth and beyond \cite{Pospergelis1969, Wolstencroft1974, Wolstencroft2004}. Recently, it has been shown that the induced fractional circular polarization by phototrophic organisms can be measured successfully in detail \cite{Sparks2009a, Sparks2009, Patty2017} and is comparable to circular dichroism measurements \cite{Patty2017}. Unlike linear spectropolarimetry, circular spectropolarimetric measurements still contain the structural information resulting from the chiral molecular systems. As such, scattered circular polarization might prove to be both a unique remotely applicable tool for vegetation monitoring on Earth as well as a powerful remotely accessible means of detecting the unambiguous presence of extraterrestrial life. 

Nonetheless, relatively few \textit{in vivo} induced circular polarization studies on phototrophic organisms are available. We previously showed that the amount of induced circular polarization of unpolarized light is equivalent to the differential absorbance of incident circularly polarized light for \textit{in vivo} transmission measurements on leaves \cite{Patty2017}. These results are evidence for at least a general cross sectional isotropy in the fractional circular polarizing/absorbing component. Little, however, is known about the possible spatial variation in the polarizing components of leaves which can offer more information about the origin of the polarization signals. 

Depending on the area of the leaf that is measured, such spatial variations might lead to inaccuracies if the molecular architecture is investigated. This is especially important for \textit{in vivo} measurements on leaves carried out using commercial dichrographs (due to the relatively small area of measurement) and it might also be important to consider when scaling up fractional polarization measurements to remote sensing applications. 

The typical circular polarization signal observed from chloroplasts is the result of the superposition of two relatively independent signals resulting from different chiral macrodomains \cite{Finzi1989}. These result in bands of opposite sign that do not have the exact same spectral shape and thus do not cancel each other out completely. The existence of these macrodomains was first demonstrated using differential circular polarization scattering \cite{Garab1988a} and the different domains were later imaged using differential polarization microscopy showing separately the positive and negative bands \cite{Finzi1989, Garab1991}. While both positive and negative signals prevail in the image averages over the whole membrane (thus including multiple macrodomains), the circular polarization spectrum is heavily influenced by the alignment of the chloroplasts \cite{Garab1991, Garab1988b, Miloslavina2012}. Local alignments of the chloroplasts might therefore affect the spatial variation in circular polarization and thus overall the signal on a leaf and canopy scale.

In the present study we will investigate the spatial components of polarization in vegetation using imaging Mueller matrix polarimetry in transmission in order to get more insight into the polarizing and depolarizing components of vegetation leaves. Various measurements on cultivated maize and maple leaves were taken within the relevant wavelength range (650 nm to 710 nm) of the vegetation absorption band in the red. We show that these measurements improve our understanding of the signals obtained on whole leaves and ultimately aid in interpreting the signals in vegetation remote sensing using circularly polarized light.

\section*{Materials and Methods}

\subsection*{Sample preparation}
Maize (\textit{Zea mays}) was grown in the laboratory of Colleen Doherty, Department of Molecular and Structural Biochemistry, North Carolina State University. The wild types we used were N78S and N74R. No differences in their growth features (V3) were observed during the measurements. The plants were cultivated in sand at a 16h/8h light-dark regime (at a photon flux density of 600 $\mu$mol m$^{-2}$ $s^{-1}$ photosynthetically active radiation (400 nm to 700 nm)) at room temperature. All spectroscopic measurements on the maize leaves were carried out with the leaves still attached to the plant. Maple (\textit{Acer rubrum}) leaves were collected in November from trees growing at the Centennial Campus, North Carolina State University in Raleigh. In order to prevent dehydration, the petioles or stems of the leaves were placed in water after collection and during the measurement.

\subsection*{Polarization and Mueller matrix decomposition}
Polarization in general is often described in terms of the four parameters of the Stokes vector $\mathbf{S}$. With the electric field vectors $E_{x}$ in the x direction ($0^{\circ}$) and $E_{y}$ in the y direction ($90^{\circ}$), the Stokes vector is given by: 

\begin{equation}
\mathbf{S}=
\begin{pmatrix}
    I\\
    Q\\
    U\\
    V\\
\end{pmatrix}=
\begin{pmatrix}
    \left\langle E^{}_{x}E^{*}_{x} + E^{}_{y}E^{*}_{y}\right\rangle\\
    \left\langle E^{}_{x}E^{*}_{x} - E^{}_{y}E^{*}_{y}\right\rangle\\
    \left\langle E^{}_{x}E^{*}_{y} - E^{}_{y}E^{*}_{x}\right\rangle\\
    i\left\langle E^{}_{x}E^{*}_{y} - E^{}_{y}E^{*}_{x}\right\rangle\\
\end{pmatrix}=
\begin{pmatrix}
    I_{0^{\circ}}+I_{90^{\circ}}\\
    I_{0^{\circ}}-I_{90^{\circ}}\\
    I_{45^{\circ}}-I_{-45^{\circ}}\\
    I_{RHC}-I_{LHC}\\
\end{pmatrix}.
\end{equation}

The Stokes parameters $I$, $Q$, $U$ and $V$ refer to intensities which thereby relate to measurable quantities. The absolute intensity is given by Stokes $I$. Stokes $Q$ and $U$ denote the differences in intensity after filtering linear polarization at perpendicular directions, where Q gives the difference between horizontal and vertical polarization and U gives the difference in linear polarization but with a 45$^{\circ}$ offset. Finally, $V$ gives the difference between right-handed and left-handed circularly polarized light. If we know the absolute intensity $I$, the polarization state can be completely described by the normalized quantities $Q/I$, $U/I$ and $V/I$. $I_{0^{\circ}}, I_{90^{\circ}}, I_{45^{\circ}}$ and $I_{-45^{\circ}}$ are the intensities oriented in the planes perpendicular to the propagation axis and $I_{LHC}$ and $I_{RHC}$ are, respectively, the intensities of right- and left-handed circularly polarized light.

Furthermore, in the Stokes formalism, any optical element can be described by the $4 \times 4$ Mueller matrix $\mathbf{M}$:

\begin{equation}
\mathbf{S}_{\mathrm{out}}=\mathbf{M}\mathbf{S}_{\mathrm{in}}=
\begin{bmatrix}
    m_{11}&m_{12}&m_{13}&m_{14}\\
    m_{21}&m_{22}&m_{23}&m_{24}\\
    m_{31}&m_{32}&m_{33}&m_{34}\\
    m_{41}&m_{42}&m_{43}&m_{44}\\
\end{bmatrix}\cdot
\begin{pmatrix}
    I\\
    Q\\
    U\\
    V\\
\end{pmatrix}_{\mathrm{in}}.
\end{equation}

The Mueller matrix elements relate to the Stokes vector by:

\begin{equation}
\mathbf{M}=
\begin{bmatrix}
	I \rightarrow I &Q \rightarrow I &U \rightarrow I &V \rightarrow I \\
I \rightarrow Q &Q \rightarrow Q&U \rightarrow Q &V \rightarrow Q \\
I \rightarrow U &Q \rightarrow U  &U \rightarrow U&V \rightarrow U \\
	I \rightarrow V &Q \rightarrow V &U \rightarrow V &V \rightarrow V\\
\end{bmatrix}
\label{eq:StokesMM}
\end{equation}

Any set of optical elements in a system can be described by a total system matrix, the product of the multiplication of the n individual elements: $\mathbf{M}=\mathbf{M}_{n}\mathbf{M}_{n-1}\ldots\mathbf{M}_{2}\mathbf{M}_{1}$. In the case of depolarizing samples such as leaves, using polar decomposition, the experimental $\mathbf{M}$ can be further decomposed into the product of a depolarizer Mueller matrix $\mathbf{M}_{\mathrm{\Delta}}$, a retarder Mueller matrix $\mathbf{M}_{\mathrm{R}}$ and a diattenuator Mueller matrix $\mathbf{M}_{\mathrm{\mathrm{D}}}$. These matrices do not commute and the result thus depends on the order of multiplication \cite{Morio2004, Lu1996, Ossikovski2007}. As there were no significant differences between illuminating a maize leaf's adaxial or abaxial side (i.e., the upper side or the under side), we have used the polar decomposition as described by Lu and Chipman \cite{Lu1996}:

\begin{equation}
\mathbf{M}=\mathbf{M}_{\mathrm{\Delta}}\mathbf{M}_{\mathrm{R}}\mathbf{M}_{\mathrm{D}}.
\end{equation}

The depolarization, and the retardance, diattenuation and their orientation can then readily be determined. The diattenuation vector $\mathbf{D}$ is given by:
\begin{equation}
\mathbf{D}\equiv D\hat{D} \equiv
\begin{pmatrix}
D_\mathrm{H} \\
D_\mathrm{45}\\
D_\mathrm{C}\\
\end{pmatrix},
\end{equation}
where $\hat{D} = \frac{\mathbf{D}}{\left| \mathbf{D}  \right|}$ and $D_{\mathrm{H}}$ is the horizontal linear diattenuation, $D_{\mathrm{45}}$ the $45^\circ$ linear diattenuation and $D_{\mathrm{C}}$ the circular diattenuation. The direction of $D$ is defined to be along the eigenpolarization with larger transmittance $(1,\hat{D}^{T})^{T}$.

The diattenuation $D$ can be defined as:
\begin{equation}
D=\left|\mathbf{D}\right| =\sqrt{D_{\mathrm{H}}^{2}+D^{2}_{\mathrm{45}}+D^{2}_{\mathrm{C}}}.
\end{equation}

It follows that:
\begin{equation}
D_{\mathrm{H}}=\frac{m_{12}}{m_{11}},D_{\mathrm{45}}=\frac{m_{13}}{m_{11}},D_{\mathrm{C}}=\frac{m_{14}}{m_{11}}.
\end{equation}
The linear diattenuation $D_{\mathrm{L}}$ is defined as:
\begin{equation}
D_{\mathrm{L}}=\sqrt{D_{\mathrm{H}}^{2}+D^{2}_{\mathrm{45}}}.
\end{equation}
The diattenuation Mueller matrix can be described by:
\begin{equation}
\mathbf{M}_{\mathrm{D}}=
\begin{bmatrix}
1 & \mathbf{D}^{\mathrm{T}}\\
\mathbf{D} & \mathbf{m_{\mathrm{D}}}\\
\end{bmatrix},
\end{equation}
with $\mathbf{m_{\mathrm{D}}}$ given by:
\begin{equation}
\mathbf{m}_{\mathrm{D}}=
\begin{bmatrix}
a+bm_{12}^2&bm_{12}m_{13}&bm_{12}m_{14}\\
bm_{13}m_{12}& a+bm_{13}^2& bm_{13}m_{14}\\
bm_{14}m_{12}& bm_{14}m_{13}& a+bm_{14}^2\\
\end{bmatrix},
a=\sqrt{1-D^2}, b=\frac{1-\sqrt{1-D^2}}{D}
\end{equation}

Similarly, the polarizance $P$, which is the polarization of unpolarized incident light, can be defined as:
\begin{equation}
P=\left|\mathbf{P}\right| =\sqrt{P_{\mathrm{H}}^{2}+P^{2}_{\mathrm{45}}+P^{2}_{\mathrm{C}}},
\end{equation}
where the diattenuation is given by the first row of $\mathbf{M}$, the polarizance is given by the first column of $\mathbf{M}$. It thus follows that:
\begin{equation}
\mathbf{P}\equiv
\begin{pmatrix}
P_\mathrm{H} \\
P_\mathrm{45}\\
P_\mathrm{C}\\
\end{pmatrix},
P_{\mathrm{H}}=\frac{m_{21}}{m_{11}},P_{\mathrm{45}}=\frac{m_{31}}{m_{11}},P_{\mathrm{C}}=\frac{m_{41}}{m_{11}}.
\end{equation}
We can then define:
\begin{equation}
\mathbf{M^\prime}\equiv\mathbf{M}\mathbf{M}_{\mathrm{D}}^{-1}=\mathbf{M}_{\Delta}\mathbf{M}_{\mathrm{R}}, \mathbf{m^\prime}=\mathbf{m}_{\Delta}\mathbf{m}_{\mathrm{R}}
\end{equation}
where $\mathbf{M^{\prime}}$ and its submatrix $\mathbf{m^\prime}$ have no diattenuation but are also not a pure retarder because of the depolarization. The depolarization $\Delta$ can be defined as:
\begin{equation}
\Delta=1-\frac{\left| \tr(\mathbf{M}_{\mathrm{\Delta}}) \right|}{3},\qquad 0\leq \Delta \leq1,
\end{equation}
where $\tr(\mathbf{M}_{\mathrm{\Delta}})$ is the sum of the main diagonal of  $\mathbf{M}_{\mathrm{\Delta}}$. $\mathbf{M}_{\mathrm{\Delta}}$ can be given by:
\begin{equation}
\mathbf{M}_{\mathrm{\Delta}}=
\begin{bmatrix}
1 &\mathbf{0}\\
\frac{\mathbf{P}-\mathbf{mD}}{1-D^2}&\mathbf{m}_\Delta\\
\end{bmatrix},
\end{equation}
and $\mathbf{m}_\Delta$ can be obtained by:
\begin{align}
\mathbf{m}_\Delta = \pm [\mathbf{m}^\prime(\mathbf{m}^\prime)^T + (\sqrt{\lambda_1\lambda_2} + \sqrt{\lambda_2\lambda_3} + \sqrt{\lambda_3\lambda_1})\mathbf{I}]^{-1}\nonumber\\
 \times [\sqrt{\lambda_1}+\sqrt{\lambda_2}+\sqrt{\lambda_3}\mathbf{m}^\prime (\mathbf{m}^\prime)^T+\sqrt{\lambda_1\lambda_2\lambda_3}],
\end{align}
where $\lambda_1,\lambda_2,\lambda_3$ are the eigenvalues of $\mathbf{m}_\Delta$. The sign depends on the determinant of $\mathbf{m}^\prime$; when the determinant is negative the sign is negative and vice versa. 
The linear depolarization $\Delta_{\mathrm{L}}$ is then given by:
\begin{equation}
\Delta_{\mathrm{L}}=1-\frac{\left| m_{\mathrm{\Delta}(11)} \right|+\left| m_{\mathrm{\Delta}(22)} \right|}{2},
\end{equation}
and the circular depolarization $\Delta_{\mathrm{C}}$ by:
\begin{equation}
\Delta_{\mathrm{C}}=1-\left| m_{\mathrm{\Delta}(33)} \right|.
\end{equation}

The retardance describes a rotation on the sphere of Poincar\'{e} and the retardance Mueller matrix $\mathbf{M_{\mathrm{R}}}$ can by described by:
\begin{equation}
\mathbf{M_{\mathrm{R}}}=
\begin{bmatrix}
1&0\\
0&\mathbf{m_{\mathrm{R}}}
\end{bmatrix},
\end{equation}
which can be obtained by:
\begin{equation}
\mathbf{M_{\mathrm{R}}}=
\mathbf{M_{\mathrm{\Delta}}}^{-1}\mathbf{M}\mathbf{M_{\mathrm{D}}}^{-1}\\
\end{equation}
The retardance vector and its fast axis $\textbf{R}$ can be defined as:
\begin{equation}
\textbf{R} \equiv R\hat{R}=
\begin{pmatrix}
Ra_1 \\
Ra_2\\
Ra_3\\
\end{pmatrix}
\equiv
\begin{pmatrix}
R_\mathrm{H} \\
R_\mathrm{45}\\
R_\mathrm{C}\\
\end{pmatrix},
\end{equation}
where the retardance, $R$, is the length of $\textbf{R}$, $\hat{R}$ is the unit vector, $R_\mathrm{H}$ is the horizontal linear retardance, $R_\mathrm{45}$ the $45^\circ$ linear retardance and $R_\mathrm{C}$ the circular retardance. 
The length of $\textbf{R}$ is given by:
\begin{equation}
R=\arccos\left[\frac{\tr(\mathbf{M}_{\mathrm{R}})}{2}-1\right],
\end{equation}
where $\tr(\mathbf{M}_{\mathrm{R}})$ is the sum of the main diagonal of $\mathbf{M}_{\mathrm{R}}$) with a fast-axis orientation defined by:
\begin{align}
\hat{R}=
\begin{pmatrix}
a_\mathrm{1} \\
a_\mathrm{2}\\
a_\mathrm{3}\\
\end{pmatrix},
a_{i}=\frac{1}{2\sin R} \sum_{j,k=1}^{3} \varepsilon_{ijk}(\mathbf{m}_{\mathrm{R}})_{jk},
\end{align}
following:
\begin{align}
(\mathbf{m}_{\mathrm{R}})_{ij}=\delta_{ij} \cos R + a_{i}a_{j}(1-\cos R) + \sum_{k=1}^{3}\varepsilon_{ijk}a_{k} \sin R, \nonumber\\
i,j=1,2,3.
\end{align}
where $\varepsilon_{ijk}$ is the Levi-Civit\'a permutation symbol, $\mathbf{m}_{\mathrm{R}}$ is a 3x3 submatrix of $\mathbf{M}_{\mathrm{R}}$ excluding the first row and column and $\delta_{ij}$ is the Kronecker delta. 
\subsection*{Mueller matrix polarimeter}

\begin{figure*}[!htb]
  \centering
  \includegraphics[width=\textwidth]{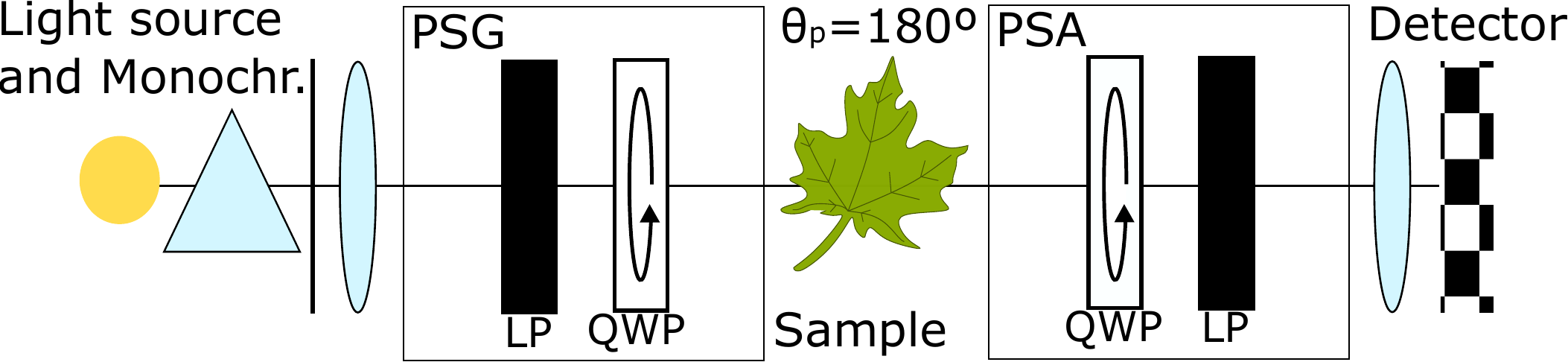}
  \caption{Schematic representation of the rotating retarder Mueller matrix ellipsometer setup in transmission, where PSG = polarization state generator, PSA = polarization state analyzer, LP = linear polarizer and QWP = quarter waveplate.}
  \label{fig:MM}
\end{figure*}

\begin{figure*}[ht]

	\centering
	\includegraphics[width=1.5\textwidth,angle =90 ]{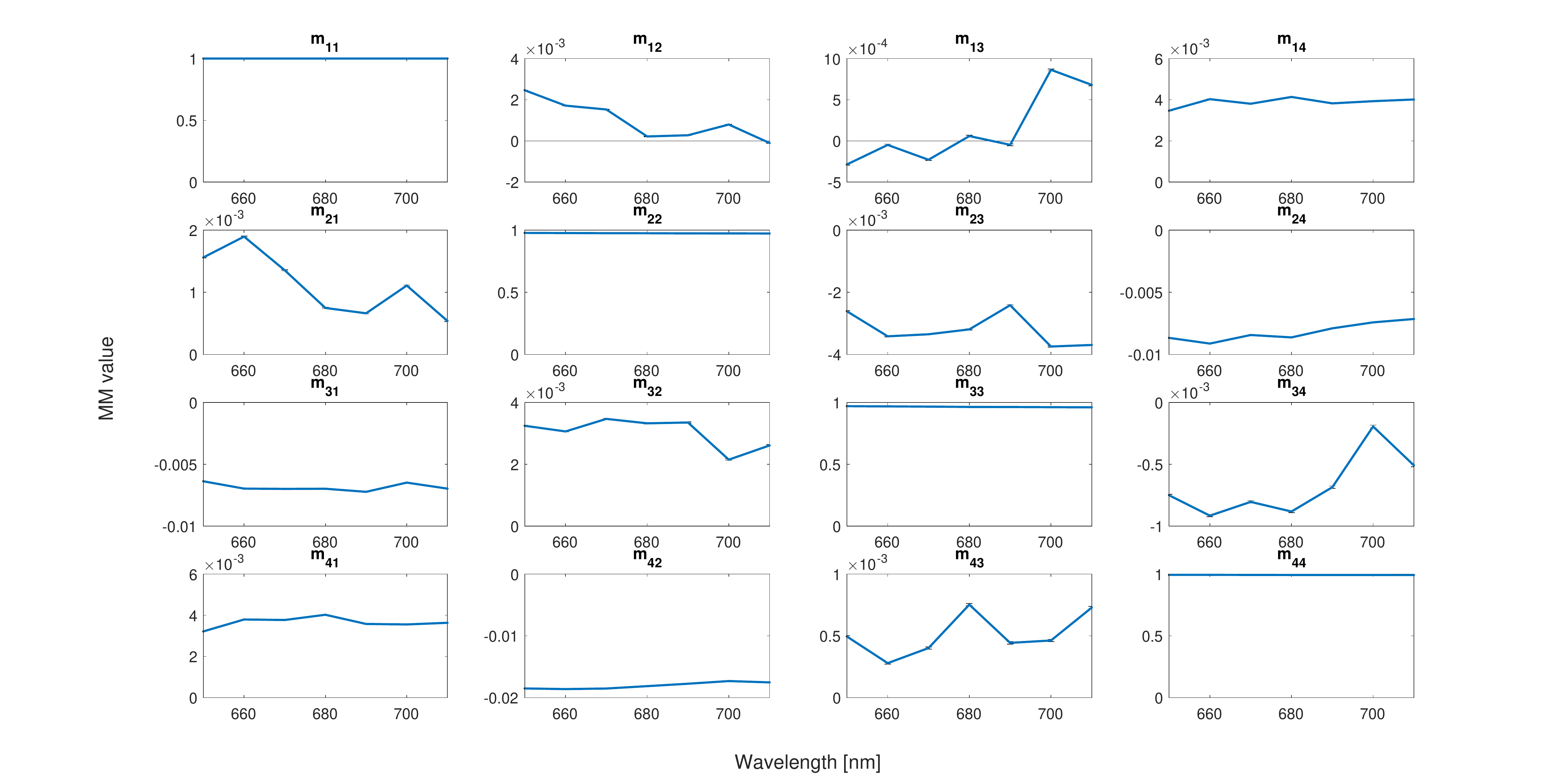}
	\caption{Wavelength dependence of the normalized Mueller matrix of an empty system (n=1). Error bars denote the standard error, but are often smaller than the graph's linewidth.}
	\label{fig:Empty}
\end{figure*}

The imaging Mueller matrix polarimeter was built by the Optical Sensing Lab (North Carolina State University). A diagram of the setup is presented in Figure \ref{fig:MM} and the wavelength dependency for the Mueller matrix elements of an empty system is given in Figure \ref{fig:Empty}. All measurements were carried out in transmission. The system was additionally verified in reflectance using a BK7 glass block and BK7 right angled prism to verify the elements relating to respectively the diattenuation and retardance. The polarimeter is based on the commonly used dual-rotating-retarder configuration as first described by Azzam \cite{Azzam1978}. To generate the polarization states, a white LED optical source, which was selected due to the high stability over time (MBB1L3, Thorlabs, USA)\footnote{Any mention of commercial products within this paper is for information only; it does not imply recommendation or endorsement by the authors or their affiliated institutions.}, was placed in front of a collimator and a monochromator with 8 nm FWHM resolution (Micro-HR, Horiba, Japan), which were followed by a polarization state generator (PSG). Hereafter the light interacted with the sample which was followed by the polarization state analyzer (PSA) for the analysis of the polarization state. A 50-mm focal length objective ($f/1.4$, AF Nikkor, Nikon, Japan) then focused the light onto a 1.2 million pixel CCD with a total spatial resolution of less than 0.1 mm per pixel (Manta G125-B, Allied Vision, Germany). Both the PSG and the PSA consisted of a fixed linear polarizer (LP) (LPVISE200, Thorlabs, USA) and a rotating quarter-wave plate (QWP) (AQWP3, Bolder Vision Optik, USA) with a retardance within 0.245 to 0.25 for the investigated wavelengths. The rotating retarders were mounted on a rotation stage (NR360S, Thorlabs, USA) driven by a stepper motor controller (BSC202, Thorlabs, USA). 

\subsection*{Data acquisition}
The polarimeter was designed to take 37 measurements for every single Mueller matrix. Obtaining a single Mueller matrix took approximately 7 minutes. The retarders rotate harmonically by a 1:5 ratio \cite{Smith2002}; per measurement the PSG QWP rotates stepwise from 0 to 180 degrees in 5 degrees increments while the PSA QWP rotates stepwise from 0 to 900 degrees in 25 degrees increments, thus resulting in different temporal modulations. The measured Stokes vector is then given by: \cite{Goldstein1992}:

\begin{equation}
\textbf{S}_{\mathrm{out}}=\mathbf{A}\mathbf{M}_{\mathrm{sample}}\mathbf{G}\textbf{S}_{\mathrm{in}},
\end{equation}
where $\mathbf{A}$ is the Mueller matrix of the PSA ($\mathbf{A}=\mathbf{M}_{\mathrm{LP}}\mathbf{M}_{\mathrm{QWP}}$), $\mathbf{G}$ the Mueller matrix of the PSG ($\mathbf{G}=\mathbf{M}_{\mathrm{QWP}}\mathbf{M}_{\mathrm{LP}}$) and $\textbf{S}_{\mathrm{in}}$ the Stokes vector of the incident light. As only the intensity is measured:

\begin{equation}
I=c\mathbf{A}\mathbf{M}_{\mathrm{sample}}\mathbf{S_{\mathrm{G}}},
\end{equation}
where c is the proportionality constant from the absolute intensity and $\mathbf{S_{\mathrm{G}}}=\mathbf{G}\textbf{S}_{\mathrm{in}}$, this can be reduced to:
\begin{equation}
I=c\sum_{i,j=1}^{4} \mu_{ij}m_{ij},
\end{equation}
where $\mu_{i,j}=a_{i}p_{j}$, with $a_{i}$ the first row of $\mathbf{A}$ and $p_{j}$ the first column of $\mathbf{G}$. As only the first row of $\mathbf{A}$ is involved:
\begin{equation}
\begin{bmatrix}
a_{1}&a_{2}&a_{3}&a_{4}\\
.&.&.&.\\
.&.&.&.\\
.&.&.&.\\
\end{bmatrix}\cdot
\begin{bmatrix}
	m_{11}&m_{12}&m_{13}&m_{14}\\
	m_{21}&m_{22}&m_{23}&m_{24}\\
	m_{31}&m_{32}&m_{33}&m_{34}\\
	m_{41}&m_{42}&m_{43}&m_{44}\\
\end{bmatrix}\cdot
\begin{pmatrix}
P_1\\
P_2\\
P_3\\
P_4\\
\end{pmatrix}=
\begin{pmatrix}
	I\\
	.\\
	.\\
	.\\
\end{pmatrix}_{\mathrm{out}}.
\end{equation}
which upon multiplication gives: 
\begin{equation}
I=\sum_{i,j=1}^{4} \mu_{ij}m_{ij},
\end{equation}
the sample Mueller matrix can then be reconstructed by multiplying the pseudo-inverse of $\mu_{ij}$ with the measured intensities.

\subsection*{Spectropolarimetry on maple leaves}
The induced fractional circular polarimetric measurements ($m_{41}$) on maple (\textit{Acer pseudoplatanus}) leaf veins were additionally measured on TreePol, a dedicated circular spectropolarimetric instrument (see \cite{Patty2017} for a description of the instrument). The leaves (n=3) were illuminated from the adaxial side, and a circular area of $\mathrm{radius}\approx $ 0.1 cm was measured. To ensure that only light from the veins was measured the other tissue was covered with black opaque PVC tape.
Additionally, the data were compared to earlier measurements of Sparks et al. \cite{Sparks2009} on older maple leaves (n=2). These measurements were carried out on a dedicated dual photoelastic modulator (PEM) polarimeter \cite{Sparks2009, Sparks2009a}, using a scanning monochromator and a three-measurements-per-point average.

\clearpage

\section*{Results}

\subsection*{Mueller matrices}
\begin{figure*}[ht]
	\centering
	\includegraphics[width=1.5\textwidth,angle =90 ]{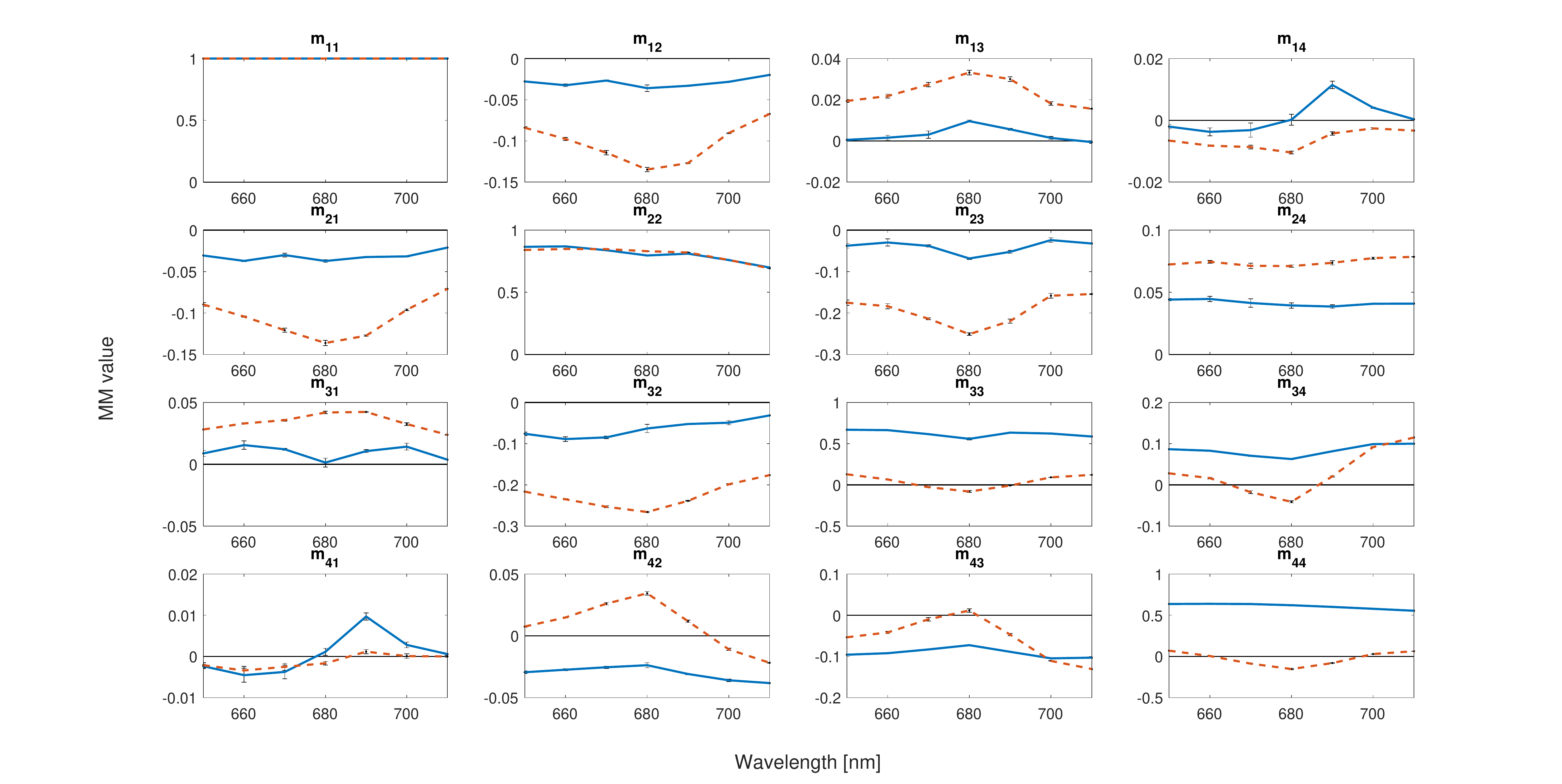}
	\caption{Wavelength dependence of the normalized Mueller matrix of \textbf{maize} leaf normal tissue (blue) and veins (orange dashed), averaged for 3 leaves. The border areas are excluded. Error bars denote the standard error.}
	\label{fig:MMmaize}
\end{figure*}

\begin{figure*}[ht]
	\centering
	\includegraphics[width=1.5\textwidth,angle =90 ]{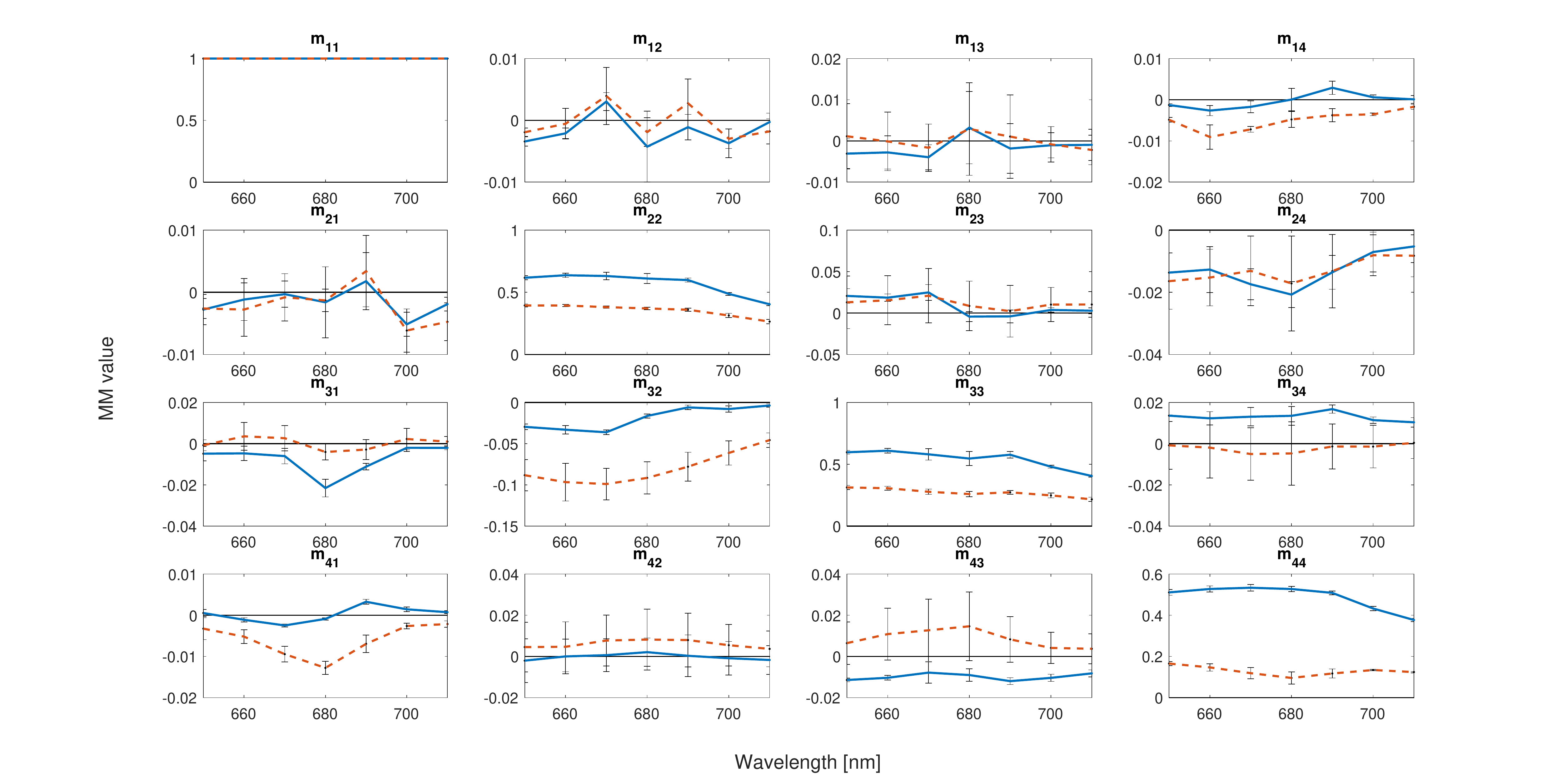}
	\caption{Wavelength dependence of the normalized Mueller matrix of \textbf{maple} leaf normal tissue (blue) and veins (orange dashed), averaged for 3 leaves. The border areas are excluded. Error bars denote the standard error.}
	\label{fig:MMmaple}
\end{figure*}

The transmission wavelength dependence of the (normalized) Mueller matrices of the normal tissue and the veins of maize leaves are shown in Figure \ref{fig:MMmaize}. Similarly, the wavelength dependence of the (normalized) Mueller matrices of the normal tissue and the veins of maple leaves as a function of wavelength are shown in Figure \ref{fig:MMmaple}. Comparing the two leaf types, roughly similar features in the individual Mueller matrix elements are visible, although with a noticeable offset in various elements. Some structure is visible in the Mueller matrix elements relating to linear polarizance ($m_{21}, m_{31}$) and dichroism ($m_{12}, m_{13}$). For maize these elements show a much stronger and gradual signal as compared to maple, which might result from the positioning of the maize leaves within the setup, which was always very similar, in combination with the parallel venation. With the exception of the elements $m_{41}$ and $m_{14}$, the signals per leaf type are generally very similar between the veins and the normal tissue, but with various offset values. The variation between the maple leaves, and thus the standard error, was much larger than that in maize leaves. These differences are likely due to the larger amount of absorbance within the maple leaves as compared to the maize leaves.

\clearpage

\subsection*{Mueller matrix elements $m_{41}$ and $m_{14}$}

\begin{figure*}[ht]
  \centering 
  \includegraphics[width=0.5\textwidth]{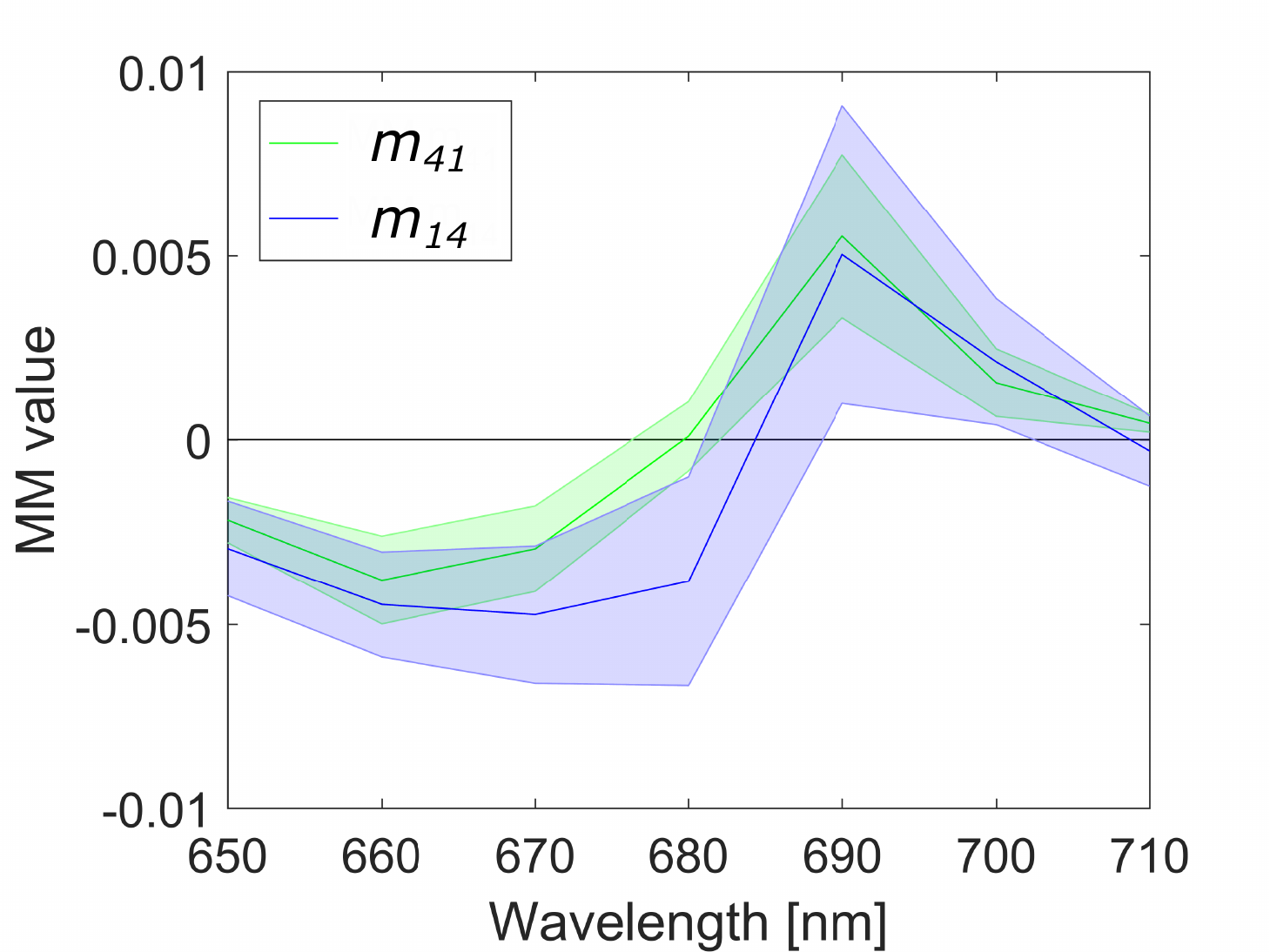}
  \caption{Wavelength dependence of the normalized Mueller matrix elements $m_{41}$ and $m_{14}$ (representing the induced fractional circular polarization and differential circular absorbance, respectively), averaged for 3 maize leaves. The shaded areas denote the standard error.}
  \label{fig:M41M14}
\end{figure*}

Figure \ref{fig:M41M14} shows that for a maize leaf the average Mueller matrix elements $m_{41}$ and $m_{14}$ are of similar shape and magnitude within the standard error. The elements $m_{41}$ and $m_{14}$ represent the induced fractional circular polarization and differential circular absorbance, respectively. The largest difference between these two elements can be found at 680 nm, which coincides with the chlorophyll maximum absorbance band and is positioned on the slope between the negative and positive peak observed in the $V/I$ signal. 

\subsection*{Spatial differences in polarization between veins and normal tissue}

\begin{figure*}[ht]
  \hspace*{-3cm}  
  \includegraphics[width=1.5\textwidth]{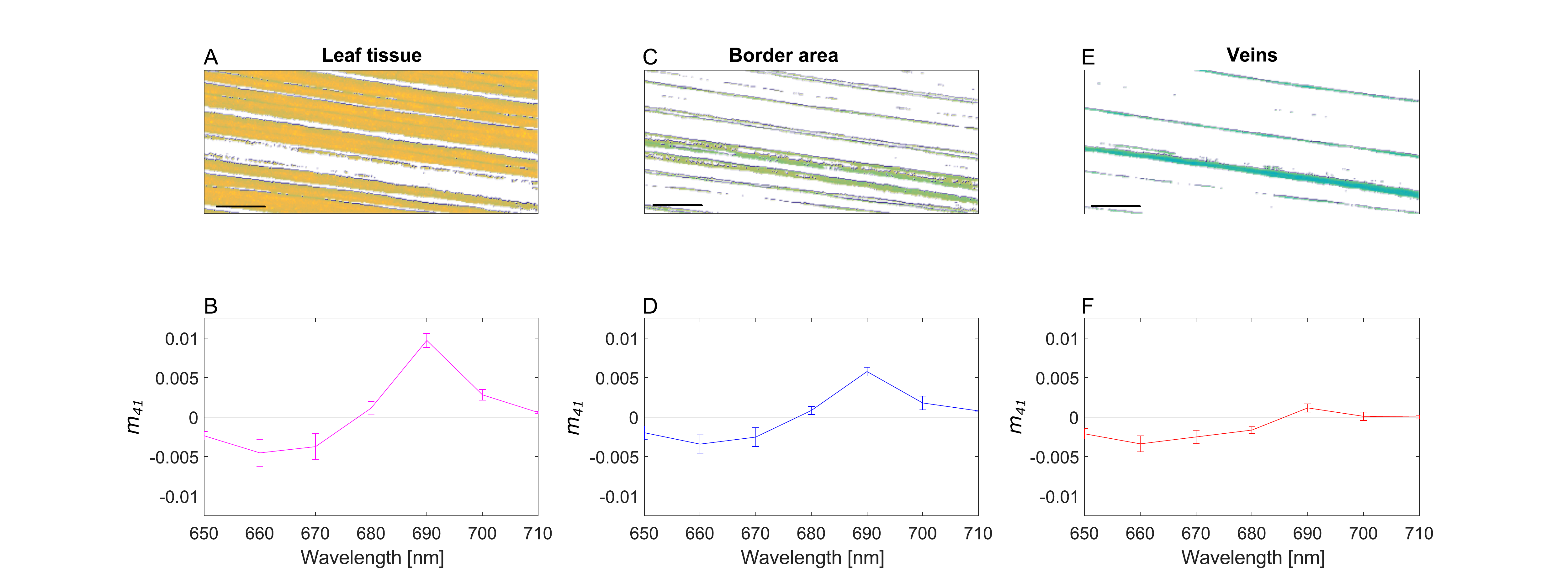}
  \caption{The different isolated spatial features of a \textbf{maize} leave (upper row) and the accompanying spectral features of Mueller matrix element $m_{41}$ (bottom row) (n=3). \textbf{A} and \textbf{B}: normal leaf tissue. \textbf{C} and \textbf{D}: the border area. \textbf{E} and \textbf{F}: the veins. Per category, the area shown in white is excluded. Scale bars in the lower left of \textbf{A},\textbf{C} and \textbf{E} are approximately 0.4 cm. Error bars denote the standard error.}
  \label{fig:M41Maize}
\end{figure*}

\begin{figure*}[ht]
  \hspace*{-3cm}
  \includegraphics[width=1.5\textwidth]{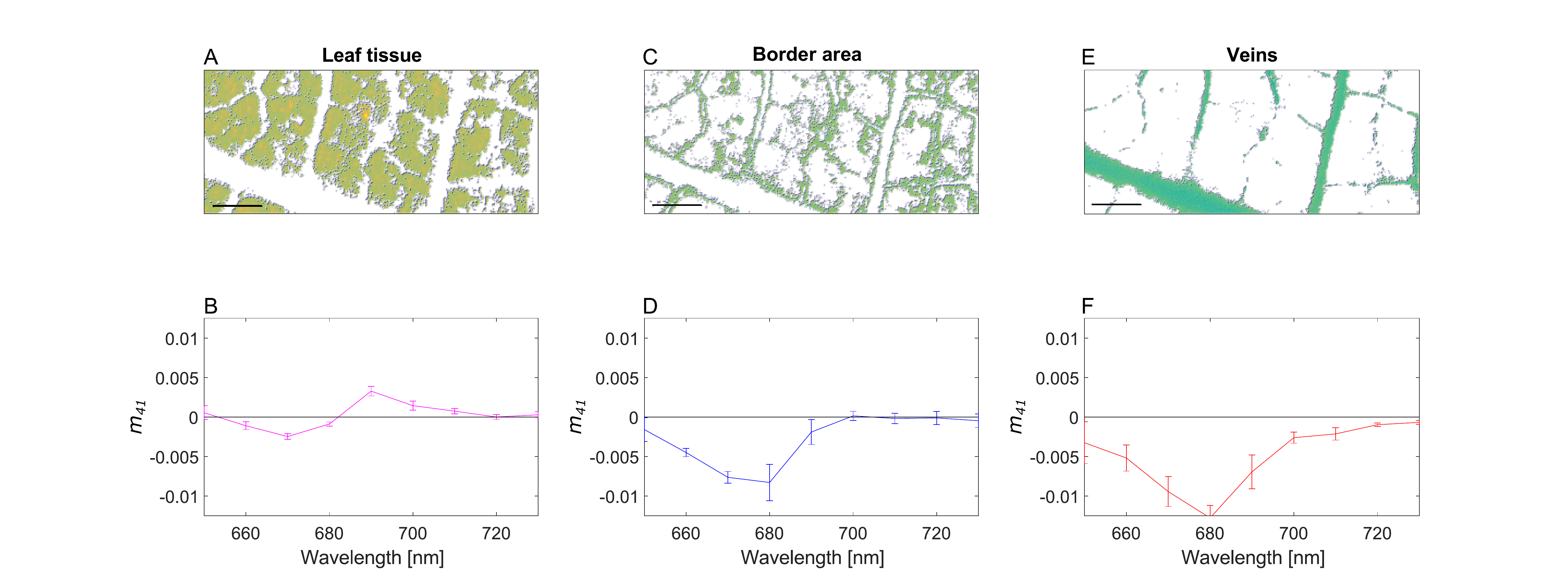}
  \caption{The different isolated spatial features of a \textbf{maple} leaf (upper row) and the accompanying spectral features of Mueller matrix element $m_{41}$ (bottom row) (n=3). \textbf{A} and \textbf{B}: normal leaf tissue. \textbf{C} and \textbf{D}: the border area. \textbf{E} and \textbf{F}: the veins. Per category, the area shown in white is excluded. Scale bars in the lower left of \textbf{A},\textbf{C} and \textbf{E} are approximately 0.4 cm. Error bars denote the standard error.}
  \label{fig:M41Maple}
\end{figure*}

As is shown in Figure \ref{fig:M41Maize} for a maize leaf and Figure \ref{fig:M41Maple} for a maple leaf, clear differences in the circular polarization features ($m_{41}$) for the selected tissue categories can be distinguished. The three categories, normal tissue, border area and veins, are discriminated on the basis of the large contrast observed in $m_{44}$, and comparing this with the total intensity ($I$) images (data not shown). In Figures \ref{fig:M41Maize} and \ref{fig:M41Maple}, the subplots \textbf{A}, \textbf{C} and \textbf{E} show the false colored image of a single measurement at 710 nm (because of the higher transmittance) of $m_{44}$ in order to highlight their spatial distributions. Excluding the white area, the average spectra (MM element $m_{41}$) of the colored areas are shown in Figures \ref{fig:M41Maize} and \ref{fig:M41Maple}: \textbf{B} \textbf{D} and \textbf{F}.

Figure \ref{fig:M41Maize} \textbf{A} and \textbf{B} show that the circular polarization per wavelength of normal maize tissue is in line with the typical signal one can expect from the measurements of thylakoid membranes \cite{Garab2009} and is similar to earlier measurements on whole leaves \cite{Toth2016, Patty2017}. In the border category, Figure \ref{fig:M41Maize} \textbf{C} and \textbf{D}, a slight decrease in the positive circular polarization band can be observed. However, looking at only the circular polarization of the veins, Figure \ref{fig:M41Maize} \textbf{E} and \textbf{F}, we can see that the positive band has almost completely disappeared while the negative band is still present and much larger in magnitude.

These differences in structural categories can be seen even more clearly for maple leaves (Figure \ref{fig:M41Maple}). Looking at the normal tissue (Figure \ref{fig:M41Maple} \textbf{A} and \textbf{B}), the shape is similar to earlier measurements on whole leaves \cite{Toth2016, Patty2017}. In the category border area (Figure \ref{fig:M41Maple} \textbf{C} and \textbf{D}) and in the veins (Figure \ref{fig:M41Maple} \textbf{E} and \textbf{F}) it is shown that the positive band is absent while the negative band has increased in signal intensity.

\begin{figure*}[ht]
  \centering
  \includegraphics[width=0.5\textwidth]{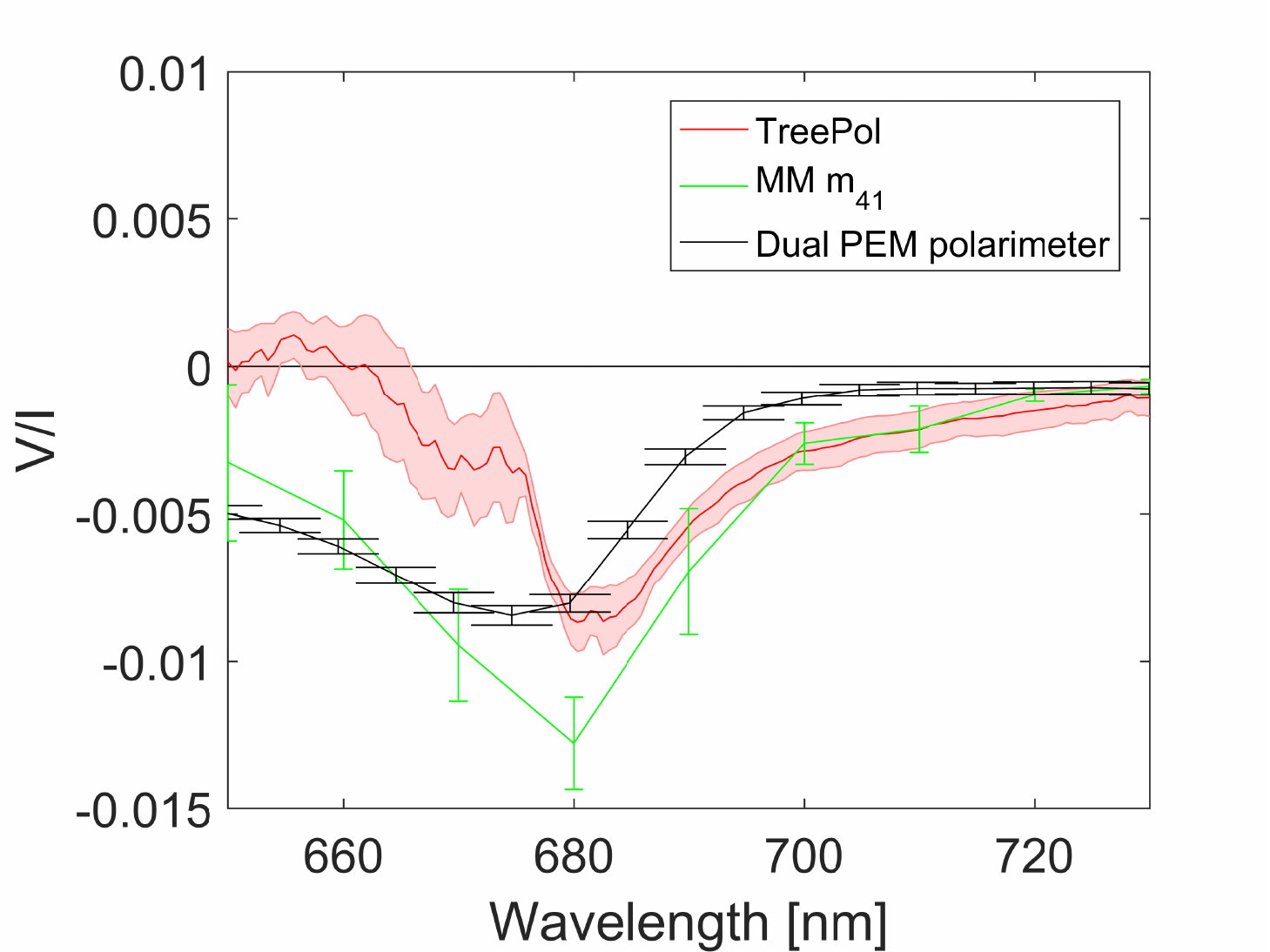}
  \caption{Transmission measurements of the veins of maple leaves carried out with TreePol, the Mueller matrix polarimeter element $m_{41}$ and of general maple leaf surfaces with the dual PEM polarimeter. Error bars and shaded area denote the standard error.}
  \label{fig:TreePol}
\end{figure*}

Circular polarization measurements specifically on veins were repeated with TreePol \cite{Patty2017}. These measurements show a signal that is similar in shape to the Mueller matrix measurements as is visible in Figure \ref{fig:TreePol} for maple leaves. TreePol has a much higher spectral resolution and shows more structure in the signal. Also shown in Figure \ref{fig:TreePol} are the results from earlier measurements on maple leaves carried out on the dual PEM polarimeter \cite{Sparks2009, Sparks2009a}. While not specifically aimed at measuring the veins, the result show a signal that is very similar to that of maple veins even though the amount of leaf tissue versus the amount of veins in these measurements is unknown.

\clearpage

\subsection*{Mueller matrix decomposition}
\begin{figure*}[ht]
	\centering
	\includegraphics[width=\textwidth]{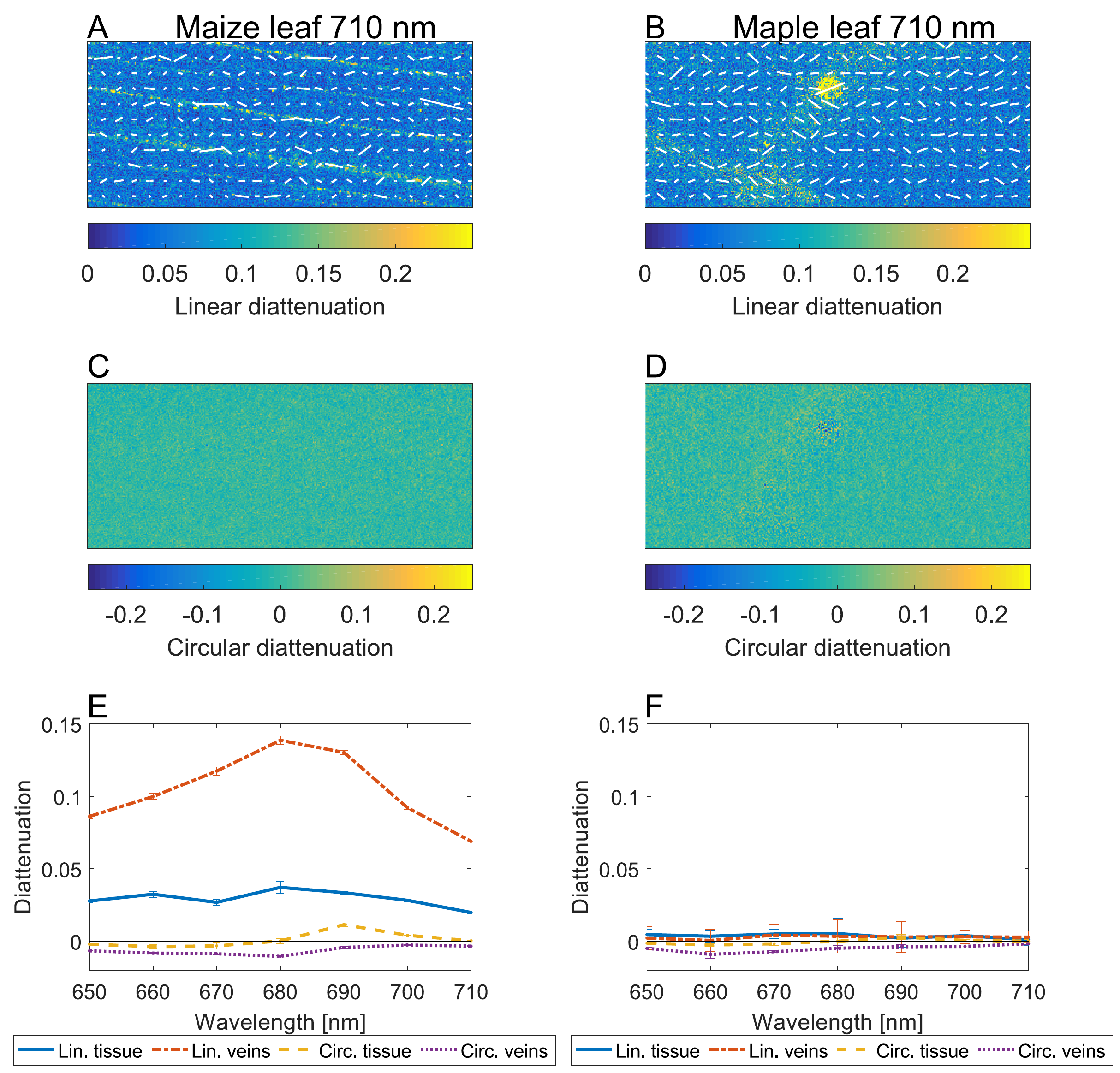}
	\caption{Spatial variations in linear diattenuation at 710 nm for \textbf{A}: maize and \textbf{B}: maple. The vectors depict the diattenuation orientation. Spatial variations in cirular diattenuation at 710 nm for \textbf{C}: maize and \textbf{D}: maple. Averaged linear and circular diattenuation over wavelength for \textbf{E} maize and \textbf{F} maple (\textit{n}=3). Error bars denote the SE.}
	\label{fig:Dia}
\end{figure*}

\begin{figure*}[ht]
	\centering
	\includegraphics[width=\textwidth]{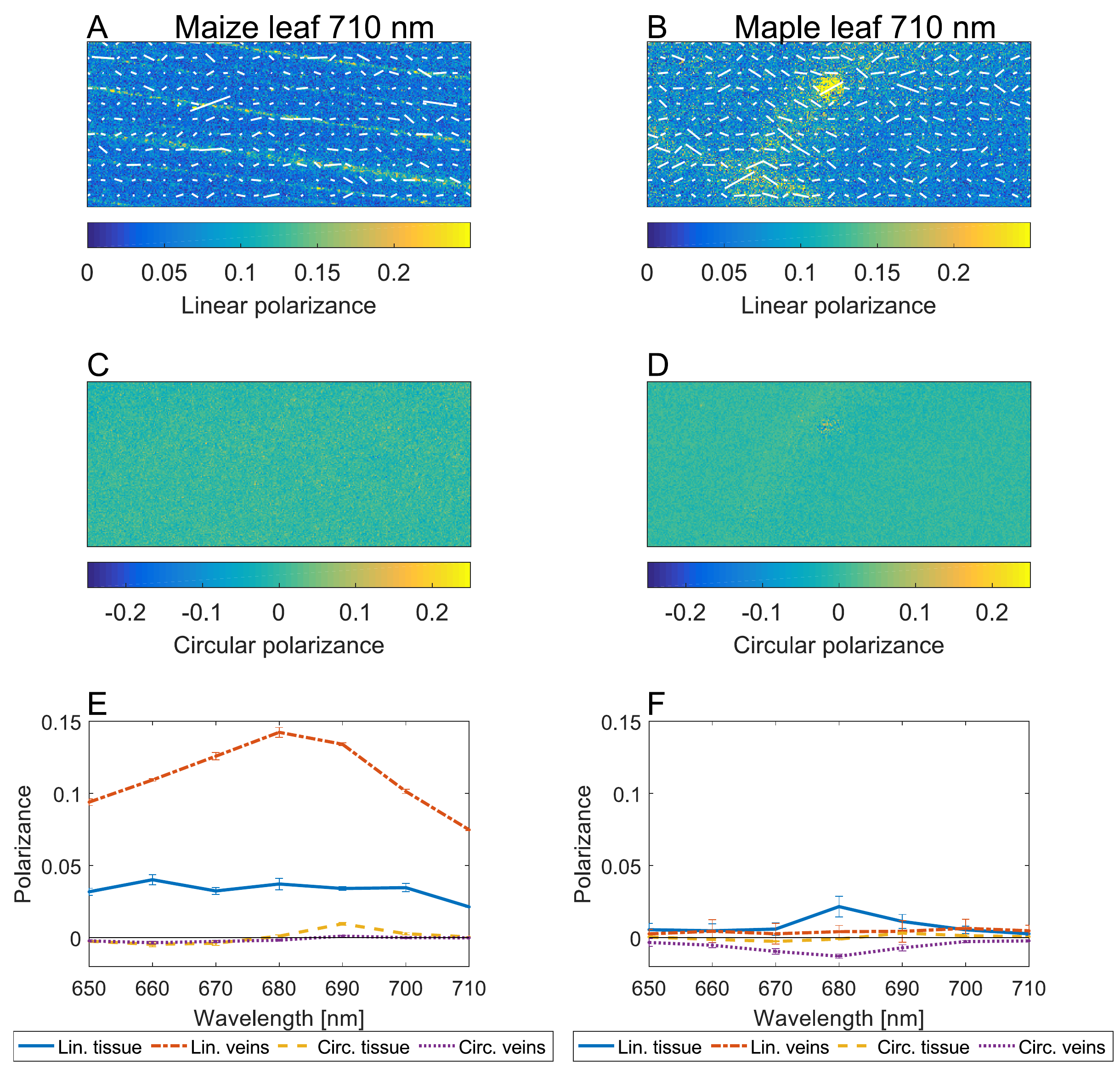}
	\caption{Spatial variations in linear polarizance at 710 nm for \textbf{A}: maize and \textbf{B}: maple. The vectors depict the polarization orientation. Spatial variations in cirular polarizance at 710 nm for \textbf{C}: maize and \textbf{D}: maple. Averaged linear and circular polarizance over wavelength for \textbf{E} maize and \textbf{F} maple (\textit{n}=3). Error bars denote the SE.}
	\label{fig:Pol}
\end{figure*}

The diattenuation for maize and maple leaves is shown in Figure \ref{fig:Dia}, where the spatial variation of the linear diattenuation at 710 nm for respectively maize and maple leaves are shown in Figure \ref{fig:Dia} \textbf{A} and \textbf{B}. In the same images the orientation of the linear diattenuation is superimposed as a vector field. In Figure \ref{fig:Dia} \textbf{C} and \textbf{D} the circular diattenuation is shown. The averages over wavelength for both linear and circular diattenuation are shown in Figure \ref{fig:Dia} \textbf{E} and \textbf{F} for respectively maize and maple. Again, similar to the associated Mueller matrix elements the linear diattenuation is observed to be larger in maize than in maple where the value is averaged out.
Similarly, the polarizance is shown in Figure \ref{fig:Pol}. For the maize leaves, the circular and linear polarizance and diattenuation are almost identical. A larger difference between those features is observed for the maple leaves, although the differences in linear and circular diattenuation and polarizance are not significant. 

\begin{figure*}[ht]
	\centering
	\includegraphics[width=\textwidth]{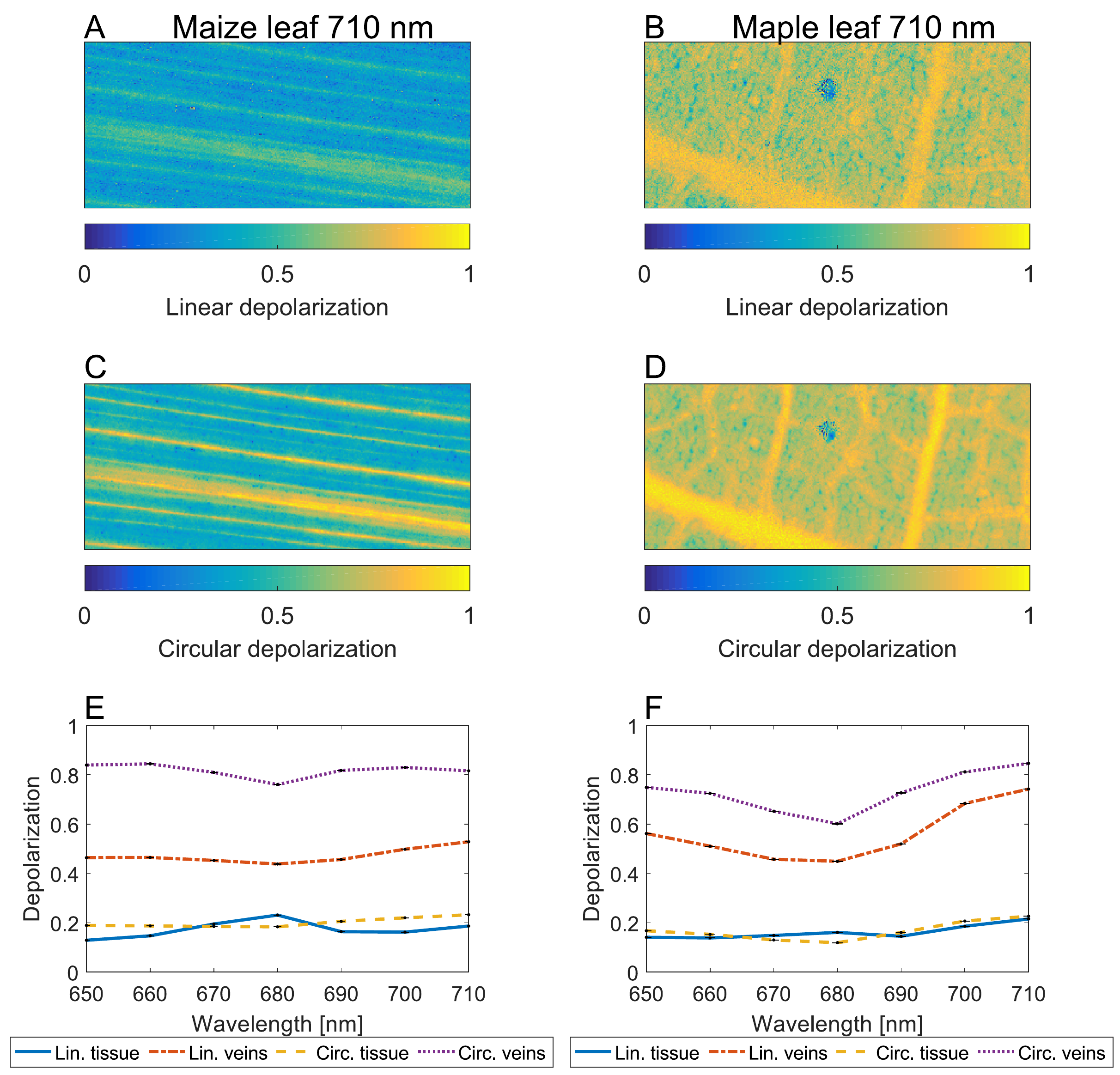}
	\caption{Spatial variations in linear depolarization at 710 nm for \textbf{A}: maize and \textbf{B}: maple. Spatial variations in cirular depolarization at 710 nm for \textbf{C}: maize and \textbf{D}: maple. Averaged linear and circular depolarization over wavelength for \textbf{E} maize and \textbf{F} maple (\textit{n}=3). Error bars denote the SE.}
	\label{fig:Delta}
\end{figure*}

The linear and circular depolarization for maize and maple leaves are shown in Figure \ref{fig:Delta}. Figure \ref{fig:Delta} \textbf{A} and \textbf{B} show the spatial variation at 710 nm of the linear depolarization for respectively maize and maple and Figure \ref{fig:Delta} \textbf{C} and \textbf{D} show the spatial variation at 710 nm of the circular depolarization for respectively maize and maple. In general, the amount of linear depolarization is much larger in the veins than in the normal tissue, which is even more pronounced for circular depolarization which is almost completely depolarized in the veins. Both the linear and circular depolarization in the veins slightly decreases in magnitude around the main chlorophyll absorbance band in the maize leaves, while this effect is larger in the maple leaves (Figure \ref{fig:Delta} \textbf{E} for maize and \textbf{F} for maple).
 
\begin{figure*}[ht]
	\centering
	\includegraphics[width=\textwidth]{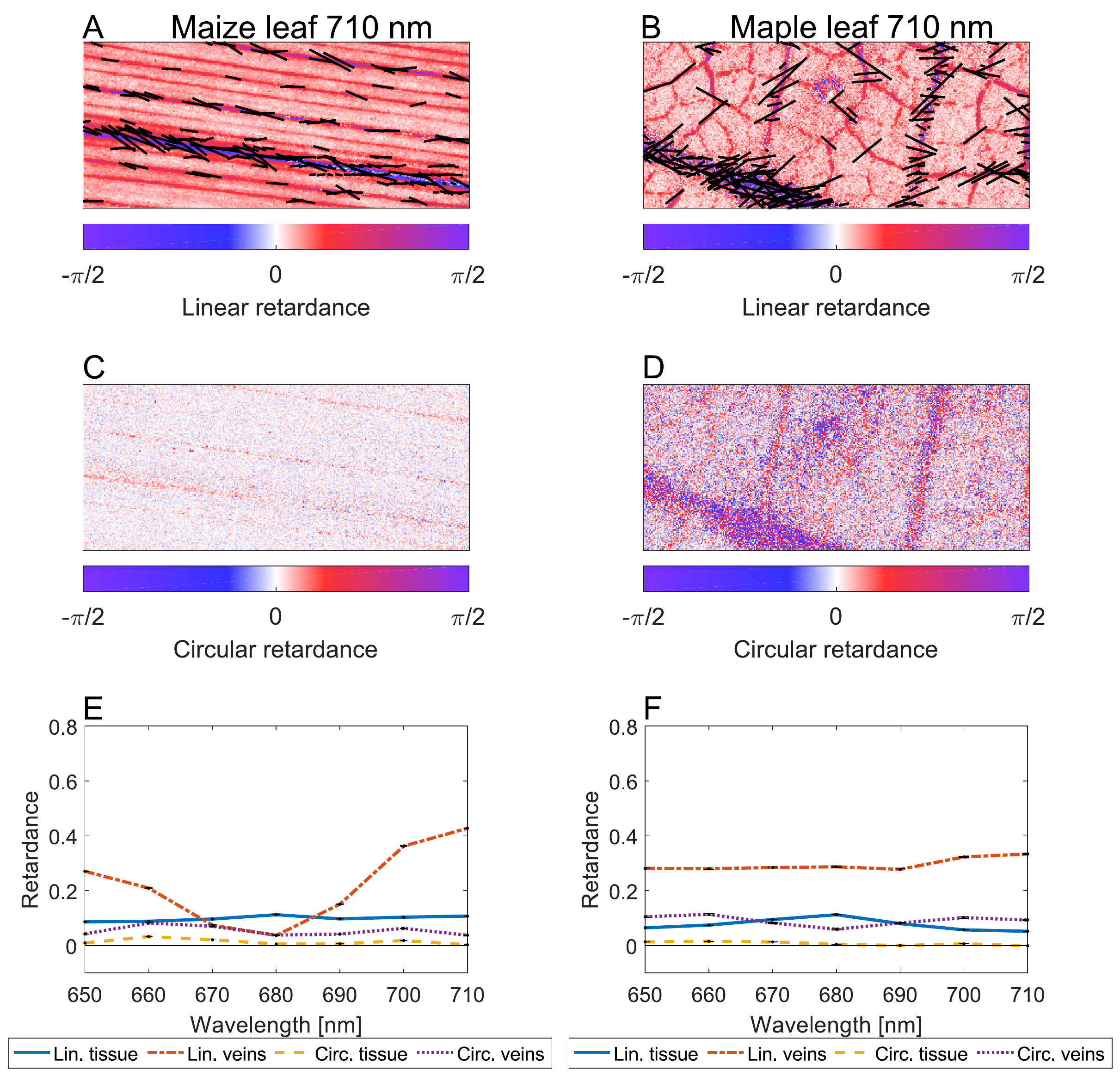}
	\caption{Spatial variations in linear retardance at 710 nm for \textbf{A}: maize and \textbf{B}: maple. The vectors depict the orientation for the retardance fast-axis and are shown only for the veins. Spatial variations in cirular retardance at 710 nm for \textbf{C}: maize and \textbf{D}: maple. Averaged linear and circular retardance over wavelength for \textbf{E} maize and \textbf{F} maple (n=3). Error bars denote the SE.}
	\label{fig:Ret}
\end{figure*}

Lastly, the retardance is shown in Figure \ref{fig:Ret}. To account for systematic offsets in the setup a rotation matrix was applied on $\mathbf{R}$. The spatial variation in linear retardance is shown for maize in Figure \ref{fig:Ret} \textbf{A} and for maple in \ref{fig:Ret} \textbf{B}. The orientation of the retardance fast-axis of the veins is shown as a vector field superimposed on the linear retardance. In both maize and maple, the retardance fast axis orientation is almost horizontal in the figure for the normal tissue, but not for the veins. Additionally, clear differences in retardance values can be observed, between the veins and the normal tissue.
The circular retardance showes more noise in the veins and as a result is slightly larger. However, no clear structures can be seen (shown in Figure \ref{fig:Ret} \textbf{C} and \textbf{D} for maize and maple respectively). No large differences in retardance are observed over wavelength as can be concluded from Figure \ref{fig:Ret} \textbf{E} (maize) and \textbf{F} (maple).

\clearpage

\section*{Discussion}
We have carried out full Mueller matrix polarimetry on various maize and maple leaves and separated the spatial features corresponding to the veins and the normal leaf tissue. While the linear diattenuation and polarizance  of maize leaves showed clear differences, these properties were averaged out in maple. This is also visible in the associated Mueller matrix elements, and we accredit the observed differences to the parallel venation of maize and the leaf orientation during the measurements which was similar for all maize leaves measured. Distinct differences between veins and normal tissue were visible in linear retardance and linear depolarization for both maize and maple. 
 
The linear polarization of vegetation has been investigated before as a remote sensing tool on Earth \cite{Peltoniemi2015, Vanderbilt1985, Vanderbilt1985a, Vanderbilt2017, Grant1993}. While indicative of leaf structural changes that can be associated with drought stress \cite{Vanderbilt2017}, the linear polarization spectral reflectance around the chlorophyll absorbance band is generally very smooth and free of structure. We did not observe the typical sharp features associated with chloroplast linear polarization \cite{Garab2009} (for Mueller matrix elements $m_{21}, m_{12}, m_{31}, m_{13}$) either, although the maize leaf veins show a smooth feature somewhat relatable to intensity. 

We observed a large difference in circular polarizance and diattenuation and the associated values of Mueller matrix elements $m_{14}, m_{41}$, between normal leaf tissue and leaf veins (Figures \ref{fig:MMmaize}, \ref{fig:MMmaple}, \ref{fig:M41Maize} and \ref{fig:M41Maple}). Normally, the spectrum of chloroplasts shows a very typical split signal around the chlorophyll absorbance band. It has been shown that this split signal is the result of the superposition of two relatively independent signals resulting from different domains \cite{Finzi1989}. The negative band has been mainly associated with the stacking of the thylakoid membranes, whereas the positive band is generally associated with the lateral organization of the chiral domains \cite{Garab1988, Cseh2000, Garab1991a}. In our measurements, the normal leaf tissue shows a typical split signal for Mueller matrix element $m_{41}$, but the veins display only a negative band. This effect was observed for both maize and maple leaves. The measurements on maple veins were repeated using TreePol, showing a roughly similar result and the overall absence of the positive band.  

The observed differences are not only due to a difference in biomolecular structures. While the bundle sheath cells in maize veins can contain chloroplasts with unstacked thylakoid membranes \cite{Faludi-Daniel1973} (which might have led to the lower signals observed in maize veins as compared to maple veins) it is known that maple does not contain similar differences between chloroplasts. Although there are definitely fewer chloroplasts and pigments around the veins, this would only lead to a smaller $V/I$ signal and would not affect the ratio between the positive and negative band.

It is on the other hand also unlikely that the effects are purely due to multiple scattering. While multiple scattering events can create circular polarization, it is not likely that scattering alone results in bands with such narrow widths since scattering polarization is usually a phenomenom leading to a very gradual wavelength dependence \cite{Martin2016}. We do assume that multiple scattering events occur near the veins, which is evident from the large amount of depolarization (see Figure \ref{fig:Delta}). The depolarization  hardly changes over wavelength (Figure \ref{fig:Delta} \textbf{E} and \textbf{F}), so it is therefore unlikely that the positive band is completely depolarized while the negative band is not.

\begin{figure*}[]
	\centering
	\includegraphics[width=\textwidth]{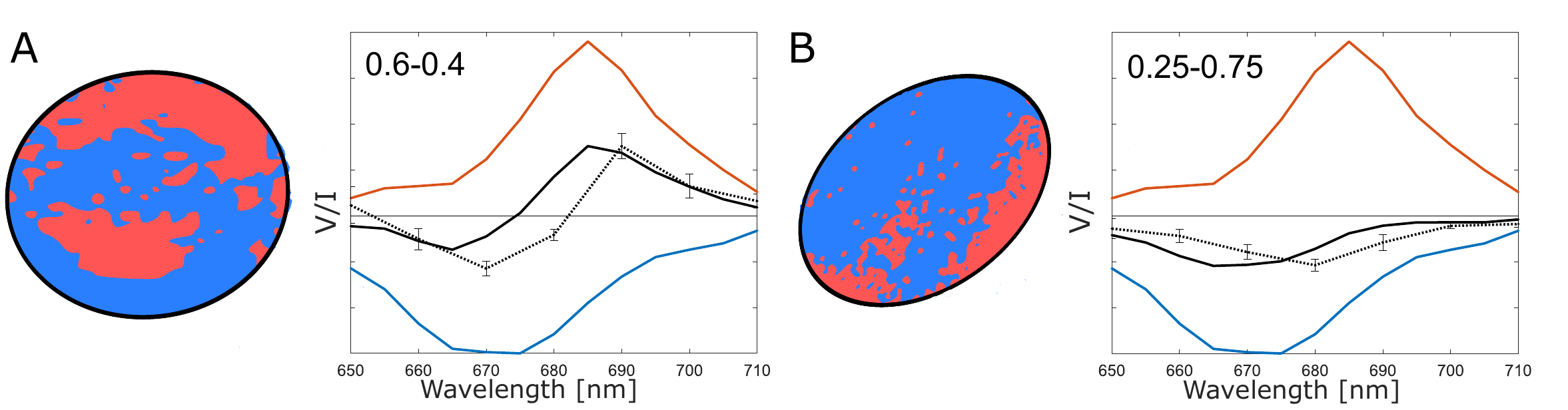}
	\caption{Representation of the chloroplast images and spectral results by Garab et al. and Finzi et al. \cite{Finzi1989,Garab1988b} including our data (normalized to superposition results). When probed using differential circular absorbance microscopy, macrodomains can be imaged having single spectral bands of opposite sign (red and blue). \textbf{A}: If the total signal (solid black line) is the superposition of $\approx 60$ $\%$ positive band and $\approx 40$ $\%$ negative band the signal is comparable to the $m_{\mathrm{41}}$ results for the normal tissue (dotted black). \textbf{B}: If the total signal (solid black line) is the superposition  $\approx 25$ $\%$ positive band and $\approx 75$ $\%$ negative band the signal is comparable to the $m_{\mathrm{41}}$ results observed for the veins (dotted black).}
	\label{fig:Dis}
	\end{figure*}

While the different macrodomains within the chloroplasts show single bands of either a positive or a negative signal, it has been reported that both bands persist in the chloroplast averages \cite{Garab1991} (a representation of these bands is shown in Figure \ref{fig:Dis}). If these bands contribute in equal amount, the superposition of the bands results in the typical split signal as observed in normal leaf tissue and randomly oriented chloroplasts in suspension. These bands do not always contribute in an equal amount, as is evident from measurements on magnetically aligned chloroplasts in suspension; the alignment of the chloroplast, be it face-aligned or edge-aligned, results in different signals \cite{Garab1991, Garab1988b, Miloslavina2012}. It should also be noted that a spectral difference was observed comparing both the negative and positive peak from the different domains in either (magnetically aligned) face or edge-aligned chloroplasts \cite{Finzi1989}. Possibly, the apparent existence of four bands is the result of superposition of two bands still persisting in the measurements on the localized 'islands', although the difference in face and edge-aligned measurements might be indicative of a spatial anisotropy in the dipole moments. 

From the images in the same study \cite{Finzi1989} it also appears that a rotational dissymetry in circular dichroism is present in the chloroplasts. As such, it seems likely that the chloroplast circular dichroism average depends on which side of the chloroplast is measured. Consequently, this feature determines the extent to which both the positive and negative bands contribute. 

We hypothesize that around the veins the chloroplasts are oriented in such a way towards the observer that the resulting signal is dominated by the negative band. In Figure \ref{fig:Dis} \textbf{A} we show that the spectral behavior of the Mueller matrix element $m_{\mathrm{41}}$ for the normal tissue of maple leaves can be reconstructed out of the two spectral bands if these have a more or less equal contribution. Figure \ref{fig:Dis} \textbf{B} shows the same results but in unequal contribution (25 $\%$ - 75 $\%$ for respectively the postive and negative bands). In this ratio the superposition is very similar to the spectral behavior of Mueller matrix element $m_{\mathrm{41}}$ observed in the veins of maple leaves. In both figures our data is red-shifted as compared to the superposition signal, but only by a few nanometers.

Additionally, it was shown that these separate signals from the macrodomains have a magnitude much larger than the superimposed signal \cite{Finzi1989}. We do not observe such large signals near the veins, which we attribute to the depolarization of the veins. The negative signal observed in maple leaf veins is still several times larger than the negative band in the split signal of leaf averages.

These findings underline that caution should be taken when scaling up small area leaf measurements to possible remote sensing applications or when evaluating measurements that use polarization modulated incident light of whole leaves to get insight into photosynthesis functioning. To illustrate this, we also included a set of measurements taken with the PEM polarimeter which were not particularly aimed at measuring the veins or the normal leaf tissue, but were taken as a general measure of leaves (see Figure \ref{fig:TreePol}). Although there is some variation between the three methods they essentially show the same pattern and any variations might be due to different contributions of the positive or negative band or slight physiological variations between the different leaves.

Importantly, these differences in circular polarization should also be considered in the evaluation of remote sensing observations itself. The measurements of whole leaves by the PEM polarimeter \cite{Sparks2009, Sparks2009a} were dominated by the negative band (Figure \ref{fig:TreePol}), but this was only the case for measurements of leaves that were collected later in the growth season. Young leaves did display the expected typical split signal (results not shown). These signals could therefore also be indicative of growth stage, depending on species, although additional measurements are required. Additionally, from an astrobiological point of view, examining the chiroptical evolution of a revolving planet might underline the presence of dynamically changing signatures of life.

Follow-up polarization microscopy studies on chloroplasts will be crucial in further evaluating these observed differences. Orientation-dependent polarization measurements using optical tweezers (see e.g. \cite{Garab2005}) should in theory allow a three-dimensional reconstruct of the chloroplast circular dichroism, which could provide a more fundamental understanding of the signal. 
\section*{Conclusion}
Using transmission imaging Mueller matrix polarimetry we have demonstrated that leaves show distinct spatial variations in linear and circular polarization characteristics as a function of wavelength. Especially in circularly polarized light we observed distinct differences in the produced fractional circular polarization and differential circular absorbance for veins and normal tissue. While the normal tissue shows the typical split sign signal comparable to circular dichroism measurements on isolated chloroplasts, the veins show only a negative band. We attribute these effects to a preferential orientation of the chloroplasts near the veins, resulting in a larger contribution of the macrodomains that display only the negative band. Although not measured in depth in this study, previously obtained data suggest a correlation with vegetation maturation. As such, these findings suggest possible applications in vegetation monitoring and may offer new prospects for the detection of extraterrestrial life by evaluating a planet's chiroptical evolution.

\section*{Acknowledgments}
This work was supported by the Planetary and Exoplanetary Science Programme (PEPSci), grant 648.001.004, of the Netherlands Organisation for Scientific Research (NWO). We acknowledge Colleen Doherty, Department of Molecular and Structural Biochemistry, North Carolina State University, for providing us with the maize samples.

\nolinenumbers
\clearpage

\bibliography{./Library/Mueller}

\bibliographystyle{ieeetr}

\end{document}